\documentclass[iop,apj]{emulateapj}

\usepackage{enumerate}

\shorttitle{Glycolaldehyde formation via the dimerisation of HCO}
\shortauthors{P.~M.~Woods et al.}

\begin{document}


\title{Glycolaldehyde formation via the \\ dimerisation of the formyl radical}


\author{Paul M. Woods\altaffilmark{1}}
\affil{Department of Physics \& Astronomy, University College London, Gower Street, London WC1E 6BT, UK.}
\email{p.woods@qub.ac.uk}

\author{Ben Slater and Zamaan Raza\altaffilmark{2}}
\affil{Department of Chemistry, University College London, 20 Gordon Street, London WC1H 0AJ, UK.}

\author{Serena Viti}
\affil{Department of Physics \& Astronomy, University College London, Gower Street, London WC1E 6BT, UK.}

\and

\author{Wendy A. Brown\altaffilmark{3} and Daren J. Burke\altaffilmark{3}}
\affil{Department of Chemistry, University College London, 20 Gordon Street, London WC1H 0AJ, UK.}


\altaffiltext{1}{present address: Astrophysics Research Centre, School of Mathematics and Physics, Queen's University Belast, University Road, Belfast, BT7 1NN, UK}
\altaffiltext{2}{present address: IMPMC, Universit{\'e} Pierre et Marie Curie, 4 Place Jussieu, Paris 75005, France}
\altaffiltext{3}{present address: Department of Chemistry, University of Sussex, Falmer, Brighton, BN1 9QJ, UK}


\begin{abstract}
Glycolaldehyde, the simplest monosaccharide sugar, has recently been
detected in low- and high-mass star-forming cores. Following on from
our previous investigation into glycolaldehyde formation
\citep{woo12}, we now consider a further mechanism for the formation
of glycolaldehyde that involves the dimerisation of the formyl
radical, HCO.  Quantum mechanical investigation of the HCO
dimerisation process upon an ice surface is predicted to be
barrierless and therefore fast. In an astrophysical context, we show
that this mechanism can be very efficient in star-forming cores. It is
limited by the availability of the formyl radical, but models suggest
that only very small amounts of CO are required to be converted to HCO
to meet the observational constraints.
\end{abstract}





\section{Introduction}

The complex organic molecule (COM) glycolaldehyde, CH$_2$OHCHO, has
been a subject of special interest among astrochemists in the last few
years. It has a number of useful qualities, including:
\begin{enumerate}[i.]
\item glycolaldehyde is distributed on compact spatial scales in
  star-forming regions, centred on protostellar cores, making it a
  tracer of early star formation;
\item it is linked to prebiotic chemistry, being involved in the
  formation of the complex sugar, ribose;
\item understanding the chemistry of glycolaldehyde and its
  relationship to its two isomers methyl formate (HCOOCH$_3$) and
  acetic (ethanoic) acid (CH$_3$COOH) gives an insight into the
  physical and chemical conditions of the star-forming core, e.g.,
  estimates of the abundance ratio between these isomers in
  Sagittarius B2(N) indicate that the structural configuration C--O--C
  belonging to methyl formate is preferred in this environment over
  the C--C--O arrangement of glycolaldehyde and acetic acid
  \citep[see][]{meh97,mil88}. Additionally, the spatial distribution
  of acetic acid is coincident with other complex species which form
  on grain surfaces, indicating that acetic acid is also the product
  of a grain surface chemistry \citep{meh97} rather than a gas-phase
  one \citep[as suggested by][for example]{hun79}.
\end{enumerate}

Glycolaldehyde has been detected towards the high-mass molecular core
\object{G31.41+0.31} \citep{bel09} and the low-mass binary
protostellar system \object{IRAS 16293-2422} \citep{jor12} on
arcsecond spatial scales. This equates to $\approx$80\,AU in the case of
\object{IRAS 16293-2422}: Solar System sizes. Its first detection in
space was towards the Galactic Centre cloud Sagittarius B2(N)
\citep{hol00}, where its distribution was widespread, in contrast to
the condensed emission regions towards the young stellar objects.

Such COMs are becoming increasingly commonly detected. In addition to
COMs detected in warm or hot cores such as \object{G31.41+0.31}, COMs
are being detected in cold, prestellar cores \citep[e.g.,][and
  references therein]{bac12,cer12}. In hot cores, it is postulated that COMs
are formed in the prestellar phase at temperatures of $\sim$10\,K, and
subsequently are evaporated into the gas phase as the forming
protostar warms its surroundings \citep{cha92}. It is not clear at
this stage to what extent COMs form in the intermediate warm-phase
(20--50\,K), where the temperatures potentially provide energy for
(large) grain-bound species to traverse the grain surface and react
before the products thermally desorb into the gas phase
\citep[e.g.,][]{gar06}. For large radicals, H atoms, which are mobile
on grain surfaces at low temperatures, are the dominant reaction
partner. However, recent experimental work \citep{fuc09} indicates
that hydrogenation (applied in this case to the hydrogenation of CO
ices) has a limited temperature window in which it is effective: above
15\,K, the end stage of hydrogenation (in this case CH$_3$OH) is
underproduced, probably due to the fact that at these warmer
temperatures H atoms desorb from surfaces more readily. H$_2$ also
desorbs at these temperatures, and since H atoms stick to H$_2$ better
than they do to CO ice, hydrogenation is inhibited. At temperatures
less than 10\,K, H atom migration is slow. At 3\,K, hydrogenation is
suppressed due to the condensation of H$_2$, meaning that no saturated
products are seen \citep{pir10}. Thus this window of 3--15\,K, which
is available in collapsing prestellar cores, but is less available in
warming protostellar cores, could potentially be crucial to the
formation of complex organic species such as glycolaldehyde.

Our understanding of the chemistry of glycolaldehyde is
developing. Several authors have suggested possible formation routes
under astrophysical conditions in the literature, and a summary is
given in \citet{woo12}. In that paper we discussed which of those
formation mechanisms are feasible in terms of reaction rates and the
availability of reactants. We focused on the isothermal collapse of a
massive molecular core, similar to that of G31.41+0.31. Using the
UCL\_CHEM chemical model \citep{vit04} we probed the efficiency of
five different mechanisms of glycolaldehyde synthesis found in the
astrophysical literature, considering a large parameter space. Of
these five mechanisms (labelled A--E), only two grain-surface routes
looked plausible under the conditions we tested:
\begin{eqnarray}
\rm A.\:CH_3OH + HCO &\longrightarrow& \rm CH_2OHCHO + H \\
\rm D.\:CH_2OH + HCO &\longrightarrow& \rm CH_2OHCHO.
\end{eqnarray}

Our focus in this paper is a possible pathway to glycolaldehyde
formation not previously considered in the literature, and we study
this method using quantum chemical techniques and astrochemical
modelling with the UCL\_CHEM code. The proposed pathway involves the
dimerisation of the formyl radical, HCO, followed by hydrogenation:
\begin{eqnarray}
\label{HCOdimerst}\rm 2HCO &\longrightarrow& \rm HOCCOH \\
\label{HCOdimermid}\rm HOCCOH + H &\longrightarrow& \rm CH_2OCHO \\
\label{HCOdimerend}\rm CH_2OCHO + H &\longrightarrow& \rm CH_2OHCHO.
\end{eqnarray}
We consider these reactions both in the gas phase and on the surface
of dust grains. From a reaction chemistry approach, this mechanism
looks promising: two reactive radicals combine to form an
intermediate, which is then hydrogenated to form glycolaldehyde. HCO
is known to exist in the gas phase in cold cores \citep{cer12} and
photon-dominated regions \citep{ger09}, for example, and is also known
to exist as an intermediate in the grain surface formation of COMs,
e.g., methanol \citep{tie82,woo02}. Hydrogenation of adsorbed species
is thought to be efficient \citep[e.g.,][]{wat03}, and thus this
reaction scheme is viable. We place these reactions into an
astrophysical context to understand their significance in
glycolaldehyde formation. We have considered both the cis- and trans-
conformers of HOCCOH, and find that the cis- conformer (leading to
cis-glycolaldehyde) is most energetically favourable in the solid
phase. Henceforth, we only consider cis-glycolaldehyde, which is the
conformer that has been detected in the interstellar medium.

\section{Calculations}
\label{sect:calcs}

\begin{deluxetable*}{lcc|cc}
%\tabletypesize{\scriptsize}
%\rotate
\tablecaption{Calculated reaction barriers on a two-bilayer crystalline
    ice slab, compared with the gas phase at the same level of
    theory. Calculations were performed at the BHandHLYP/DZVP level. \label{tab:newbarriers}}
\tablewidth{0pt}
\tablehead{
\colhead{Reaction} & \multicolumn{2}{c}{Ice-surface barrier} & \multicolumn{2}{c}{Gas-phase barrier} \\
\colhead{} & \colhead{(kJ\,mol$^{-1}$)} & \colhead{(K)} & \colhead{(kJ\,mol$^{-1}$)} & \colhead{(K)}
}
\startdata
\phantom{CH$_2$}2HCO\phantom{\,\,\,+ H} $\longrightarrow$ HOCCOH & 0.00 &  \phantom{100}0 & \phantom{1}0.00 &  \phantom{100}0  \\
\phantom{$_2$}HOCCOH + H $\longrightarrow$ CH$_2$OCHO & 9.18 & 1\,108 & 10.87 & 1\,312  \\
CH$_2$OCHO + H $\longrightarrow$ CH$_2$OHCHO & 0.00 &  \phantom{100}0  & \phantom{1}0.00 &  \phantom{100}0  
\enddata
\end{deluxetable*}

In this work we have used a combination of periodic and aperiodic
Density Functional Theory (DFT) approaches to explore the influence of
a proto-dust grain on the mechanism of glycolaldehyde formation. A 3D
periodic slab was used to model the ice surface, onto which reactants
are adsorbed. It has been assumed that most of the interstellar ice
that coats an ISM dust grain takes the form of amorphous solid water
\citep[ASW;][]{gib00}. Modelling such a high degree of local disorder
would require a large unit cell, making hybrid functional DFT
calculations prohibitively expensive. Therefore, as a first-order
approximation, to probe the influence of the substrate chemistry, we
employ a two bilayer crystalline slab containing 96 water molecules,
with a 35\,\AA\ vacuum gap. The crystalline ice phase used, Ih,
exhibits orientational disorder in the hydrogen positions. Within the
constraints of the lattice, each tetrahedrally coordinated water
molecule can have one of six different positions, but must obey the
Bernal-Fowler-Pauling ice rules, which require that each oxygen atom
has two nearest neighbour hydrogen atoms to form a water molecule, and
there must be exactly one hydrogen atom on a hydrogen bond joining two
nearest neighbour oxygen atoms. Rick and Haymet's ``move'' algorithm
\citep{Rick2003} was used to generate disordered configurations
without violating the ice rules. It has been demonstrated that proton
disorder on the ice surface has a much larger effect on surface energy
than on the bulk cohesive energy \citep[see][]{Pan2010}. Here, we
selected an approximately random surface ordering (an order parameter
of 3.3, using the notation of Pan et al.) to provide a variety of
adsorption sites \citep[see][]{Watkinspnas,Watkinsnaturem}.

The \textsc{Quickstep} module of the CP2K suite
\citep{Vandevondele2005} was used for all surface calculations since
the recently implemented Auxillary Density Matrix Method
\citep[ADMM;][]{Guidon2010} allows hybrid DFT calculations to be
completed on a similar timescale to Generalised Gradient Approximation
(GGA) calculations. \citet{Andersson2004} showed that for
astrochemical reactions, hybrid functionals are often essential to
describe barriers reasonably well. We opted to use the BHandHLYP
hybrid density functional together with Goedecker-Teter-Hutter (GTH)
pseudopotentials and double-zeta (DZVP) basis sets, a 400\,Ry plane
wave cutoff and the DFT-D3 \citep{Grimme2010} dispersion correction
with a cutoff of 10\,\AA. The BHandHLYP functional performs
significantly better than, for example, B3LYP, for calculating
reaction heats and barriers where one of the reactants is a hydrogen
atom \citep{Andersson2004}. It consistently and significantly (but not
catastrophically, like B3LYP) underestimates barriers for the gas
phase reactions; however, since we are interested in changes in the
barriers, and in the absence of more finely tuned functionals such as
M05-2x, it is a reasonable choice. Convergence tolerances were set to
a minimum energy change of 10$^{-6}$\,Ha for electronic steps, and a
maximum displacement of 10$^{-3}$\,Bohr and a maximum force of
5$\times$10$^{-5}$\,Ha/Bohr for ionic steps. The cpFIT3 auxiliary
basis set was used for the ADMM, with a Coulomb truncation radius of
5\,\AA.

The focus of our calculations in this paper can be seen in
Table~\ref{tab:newbarriers}, which shows our new pathway for the
formation of glycolaldehyde. Two of the three reactions of interest in
the mechanism are barrierless in the gas phase and they are also found
to be barrierless on the surface. Only the reaction HOCCOH + H
$\longrightarrow$ CH$_2$OCHO has a substantial barrier and we find a
very modest reduction in the barrier height of
$\sim$1.7\,kJ\,mol$^{-1}$ (200\,K) when the reaction is carried out on
the substrate. Reverse barriers for the reactions have also been
calculated, and all are sufficiently larger than the forward reaction
barrier (all reverse barriers are $>$100\,kJ\,mol$^{-1}$ (12\,000\,K)
in height), suggesting that the reverse reaction is unlikely.

Since only one step in the reaction pathway discussed here has a
substantial barrier, we consider the nature of the barrier and factors
which could influence the predicted barrier. Before addition of H to
HOCCOH can occur, two HCO molecules must combine. We calculated the
migration barrier for self-diffusion of HCO on the ice surface, by
rastering the molecule over the surface (using a fixed distance
constraint and incrementally moving the HCO radical by 0.1\AA\ along
the $b$-axis of the slab) and found that migration across the surface
encounters barriers of $<$2\,kJ\,mol$^{-1}$ ($<$240\,K), according to
the DFT model used. This barrier is sufficiently small that it will be
overcome under the low temperature conditions considered here. The
BHandHLYP estimate of the reaction barrier height of the gas phase is
around a third of the CCSD(T) estimate, hence the barrier height on
the surface is likely to be underestimated. However, we only
considered a small sample of adsorption sites on the ice surface and
it is likely that more favourable adsorption sites exist in porous
ASW, which would reduce the barrier (i.e., a stronger adsorption of
HOCCOH on the ice would activate the site on the carbon atom for an
attack by H). The absolute rate of reaction is expected to be
dominated by tunnelling and indeed the crossover temperature is
estimated to be 172\,K \citep{gillan}, confirming that tunnelling will
account for the apparent rate. A recent study of H addition to
polycyclic aromatic hydrocarbons showed that rate coefficients are
dramatically enhanced by tunnelling and indeed tunnelling becomes the
dominant mechanism at ISM temperatures \citep{Goumans2011}. Previous
studies indicate that the quantum tunnelling rate can be $>$10$^5$
faster than the classical rate at the temperature range relevant here
and for reactions with similar barrier heights. In
\citet{andersson_cpl_2011}, bimolecular addition of H to CO has a
barrier of $\sim$12\,kJ\,mol$^{-1}$ (1440\,K, in close agreement with
our value for H addition to a C centre) which was found to yield a
quantum tunnelling rate dominated value of
7$\times$10$^{-17}$\,cm$^3$\,s$^{-1}$. This rate informs our
sensitivity study in the following section.

We also note that reaction~\ref{HCOdimerst} in our mechanism, HCO +
HCO $\longrightarrow$ HOCCOH, will compete with the hydrogenation of
HCO on grain surfaces to form H$_2$CO. This hydrogenation reaction is
also barrierless, and co-incidentally forms a part of another route
towards the formation of glycolaldehyde which we discussed previously
\citep[mechanism D,][]{woo12}, and which is adopted by other authors
\citep[e.g.,][]{gar06,gar08}. We will show in a future paper that both
these pathways for the reaction of HCO (via dimerisation or via
formaldehyde and hydroxymethyl intermediates) could potentially lead
to glycolaldehyde. For the remainder of this paper, however, we turn
our attention back to the dimerisation of HCO.

\section{Chemical modelling}

In order to understand the effectiveness of this pathway to
glycolaldehyde formation, we simulate an astrophysical environment
with a chemical model. We use the UCL\_CHEM chemical model in much the
same way as in \citet{woo12}: we consider the isothermal free-fall
collapse of a prestellar core \citep[see][]{raw92} from a diffuse
medium until a density appropriate for a star-forming core is reached:
$n_\mathrm{fin}\sim$10$^7$\,cm$^{-3}$ for a high-mass core;
$n_\mathrm{fin}\sim$10$^8$\,cm$^{-3}$ for a low-mass core. These
densities are typical for the molecular regions near the centre of
pre-/proto-stellar cores. The process of the isothermal collapse of
the core we call Phase {\sc i}. As the collapse progresses in our
model, gas-phase molecules are accreted onto the surface of dust
grains, where they may undergo hydrogenation or reactions which may
lead to the formation of glycolaldehyde only (reactions
\ref{HCOdimerst}--\ref{HCOdimerend}).  Imposing this constraint means
that we take a conservative approach in terms of method, and a
generous approach in terms of the formation of glycolaldehyde.  Since
we consider only the most favourable formation of glycolaldehyde, we
derive an upper limit to abundance estimates for this particular route
of formation. Within this model, CO is hydrogenated to CH$_3$OH via
intermediates HCO and H$_2$CO, C hydrogenated to CH$_4$, N
hydrogenated to NH$_3$ and O hydrogenated to H$_2$O, etc. Thus
adsorbed HCO, which is important for our formation mechanism, can be
formed either through the hydrogenation of adsorbed CO or through the
freeze-out of gas-phase HCO.

Following the collapse, there is a \textquotedblleft
warm-up\textquotedblright\ phase, which we call Phase {\sc ii}, where
the collapse ceases and grain mantles are evaporated as the protostar
warms its surroundings. No grain-surface reactions occur in the model
in this phase, due to the large uncertainties in the treatment of the
process.

\subsection{A note on reaction rate coefficients}

We have investigated whether glycolaldehyde production via the
mechanism described in reactions~\ref{HCOdimerst}--\ref{HCOdimerend}
is efficient and significant. Reaction rates for the latter two
reactions have not been measured to our knowledge; however, rates for
the reaction of HCO with itself have been quantified experimentally in
the gas phase \citep[e.g.,][see
Table~\ref{tab:rates}]{fri02,yee68}. Three branches have been
identified:
\begin{eqnarray}
\nonumber\rm 2HCO &\longrightarrow& \rm CO + H_2CO \\
\nonumber &\longrightarrow& \rm HOCCOH\footnotemark\\
\nonumber &\longrightarrow& \rm CO + CO + H_2 
\end{eqnarray}
and measured reaction rate coefficients can be found in
Table~\ref{tab:rates}.

\footnotetext{Reported in the original paper as (CHO)$_2$.}

In the modelling of reaction~\ref{HCOdimerst}, 2HCO $\longrightarrow$
HOCCOH, we adopt a reaction rate coefficient of
2.8$\times$10$^{-13}$\,cm$^3$\,s$^{-1}$, as measured in the gas phase
by \citet{yee68}, for both gas-phase and solid-phase reactions. This
has little physical significance as a surface reaction rate, but we
use the numerical value as an equivalent. We have estimated the
diffusion barrier to be $<$240\,K, which is reasonably overcome at
10\,K, since the molecules are physisorbed $\sim$3\AA\ above the
surface.

The rate coefficients for the remaining two reactions,
\ref{HCOdimermid} and \ref{HCOdimerend}, have not been measured to our
knowledge. In the gas phase, these reactions are presumably slow
associations, and would more readily occur as protonation
reactions. Comparison with similar reactions in the UMIST and KIDA
databases show that reaction rate coefficients are vanishingly small
at 10\,K, and most lead to two products, one of which is H$_2$ (i.e.,
abstraction dominates over addition). Given that rate coefficients for
these reactions are $\sim$10$^{-11}$\,cm$^3$\,s$^{-1}$, and we expect
H-addition to be a minor channel, we adopt gas-phase rate coefficients
of 1$\times$10$^{-14}$\,cm$^3$\,s$^{-1}$ for reactions
\ref{HCOdimermid} and \ref{HCOdimerend}. The choice of gas-phase rate
coefficient for this reaction is not crucial to the results of the
model.

In the solid phase, as previously mentioned,
\citet{andersson_cpl_2011} calculated a rate coefficient of
7$\times$10$^{-17}$\,cm$^3$\,s$^{-1}$\ for H + CO at 20\,K, including
a tunnelling correction. This reaction has a classical barrier of
1\,500--1\,850\,K. Similarly, \citet{gou11} calculated a rate
coefficient of $\sim$10$^{-21}$\,cm$^3$\,s$^{-1}$\ for H + H$_2$CO at
20\,K (with a barrier of 2\,318\,K). Since HOCCOH and OCH$_2$CHO are
approximately twice as massive as H$_2$CO we can crudely expect an
increase in the rate of hydrogen addition, and we consider the range
of rate coefficients from 10$^{-16}$ to
10$^{-21}$\,cm$^3$\,s$^{-1}$\ reasonable for these reactions. To
maintain consistency we assume that on grain surfaces these two
hydrogenation reactions proceed at the same rate as the other grain
surface hydrogenation reactions previously mentioned. For these two
reactions, this rate coefficient is on the order of
10$^{-19}$\,cm$^3$\,s$^{-1}$, which falls neatly within the acceptable
range.

Reaction~\ref{HCOdimermid} is retarded by a factor $e^{-\gamma/T}$,
where $\gamma$\ is the energy barrier for the reaction, in Kelvin. As
mentioned above, reaction~\ref{HCOdimermid} is subject to an
activation energy barrier of at least 1\,312\,K in the gas phase, and
1\,108\,K on a water ice surface. Tunnelling of H atoms may be
possible through both these barriers, as discussed in
Section~\ref{sect:calcs} and above. In our modelling we experiment
with different values for the energy barriers, but maintain the
reaction rate coefficient mentioned above.

\begin{deluxetable*}{lccl}
%\tabletypesize{\scriptsize}
%\rotate
\tablecaption{Experimental reaction rates for HCO + HCO $\longrightarrow$ products. \label{tab:rates}}
\tablewidth{0pt}
\tablehead{
\colhead{Reaction} & \colhead{Rate coef. (cm$^3$\,s$^{-1}$)} & \colhead{Source} & \colhead{Reference} 
}
\startdata
2HCO $\longrightarrow$ CO + H$_2$CO & 7.5$\times$10$^{-11}$ & NIST & \citet{bag86} \\
& 6.3$\times$10$^{-11}$ & NIST & \citet{rei78} \\
& 5.0$\times$10$^{-11}$ & KIDA & undisclosed, probably \citet{bau92} \\
& 4.5$\times$10$^{-11}$ & KIDA & \citet{fri02} \\
& 3.4$\times$10$^{-11}$ & NIST & \citet{vey84} \\
& 3.0$\times$10$^{-11}$ & UDFA & probably \citet{sar84} \\
2HCO $\longrightarrow$ HOCCOH\footnotemark[1] & 5.0$\times$10$^{-11}$ & NIST & \citet[][unconfirmed]{sto85}$^\dagger$ \\
& 2.8$\times$10$^{-13}$ & -- & \citet{yee68} \\
2HCO $\longrightarrow$ 2CO + H$_2$ & 3.6$\times$10$^{-11}$ & KIDA & \citet{yee68} 
\enddata
%% Text for table notes should follow after the \enddata but before
%% the \end{deluxetable}. Make sure there is at least one \tablenotemark
%% in the table for each \tablenotetext.
\tablecomments{Sources: NIST (\url{http://kinetics.nist.gov}); UDFA
  (\url{http://udfa.net}); KIDA
  (\url{http://kida.obs.u-bordeaux1.fr}). $^\dagger$\citet{sto85}
  give this gas-phase rate coefficient in relation to the
  recombination of the HCO radical, but the reaction products are not
  stated.}
\end{deluxetable*}
%\footnotetext{Reported in the original paper as (CHO)$_2$.}
\begin{deluxetable*}{lcccccccc}
\tabletypesize{\footnotesize}
%\rotate
\tablecaption{Summary of model parameters for 25\,M$_\odot$\ models. \label{tab:models}}
\tablewidth{0pt}
\tablehead{
\colhead{\#} & \colhead{Gas-phase} & \colhead{Grain-surface} & \colhead{CO
  $\rightarrow$ HCO} & \colhead{$fr$} &
 \colhead{$t_\mathrm{fin}$} & \colhead{Non-thermal} &
\colhead{$x$(CH$_2$OHCHO)} & \colhead{$x$(CH$_2$OHCHO)} \\ \colhead{} & \colhead{barrier (K)} & \colhead{barrier
  (K)} & \colhead{(\%)} & \colhead{} &
\colhead{(yr)} & \colhead{desorption} & \colhead{Phase {\sc i}}& \colhead{Phase {\sc ii}}
}
\startdata
1 & 1\,312 & 1\,108 & \ldots & 0.10 & n.c. & \ldots & negligible & 6.0$\times$10$^{-18}$\\
2 & 1\,312 & \ldots & \ldots & 0.10 & n.c. & \ldots & 2.4$\times$10$^{-12}$ & 2.4$\times$10$^{-12}$  \\
3 & \phantom{1\,}656 & \ldots & \ldots & 0.10 & n.c. & \ldots & 2.4$\times$10$^{-12}$ & 2.4$\times$10$^{-12}$ \\ \tableline
4 & 1\,312 & \ldots & 25 & 0.10 & n.c. & \ldots & 7.9$\times$10$^{-6\phantom{1}}$ & 7.9$\times$10$^{-6\phantom{1}}$ \\
5 & 1\,312 & \ldots & 25 & 0.10 & 10$^7$ & \ldots & 1.3$\times$10$^{-5\phantom{1}}$ & 1.3$\times$10$^{-5\phantom{1}}$\\
6 & 1\,312 & \ldots & 25 & 0.10 & n.c. & H$_2$ formation & 1.1$\times$10$^{-5\phantom{1}}$ & 1.1$\times$10$^{-5\phantom{1}}$ \\
7 & 1\,312 & \ldots & 25 & 0.10 & n.c. & all & 1.1$\times$10$^{-5\phantom{1}}$ & 1.1$\times$10$^{-5\phantom{1}}$\\
8 & 1\,312 & \ldots & 25 & 0.20 & n.c. & \ldots & 7.4$\times$10$^{-6\phantom{1}}$ & 7.4$\times$10$^{-6\phantom{1}}$\\
9 & 1\,312 & \ldots & 25 & 0.05 & n.c. & \ldots & 4.5$\times$10$^{-6\phantom{1}}$ & 4.5$\times$10$^{-6\phantom{1}}$ 
\enddata
\tablecomments{
Reaction energy barriers refer to reaction~\ref{HCOdimermid}. In this
table, n.c. indicates \textquoteleft no
constraint\textquoteright\ upon $t_\mathrm{fin}$. Typically this means
that the collapse reaches $n_\mathrm{fin}$ in
$\sim$5$\times$10$^6$\,yr. $n_\mathrm{fin}$ for all models is
10$^7$\,cm$^{-3}$.  The conversion between CO and HCO happens via
grain-surface hydrogenation upon freeze-out (column 4; see text.)
Glycolaldehyde abundances are given in the solid phase for Phase {\sc
  i} and gas phase for Phase {\sc ii}.}
\end{deluxetable*}

\subsection{Modelling a high-mass (25\,M$_\odot$) core}

\object{G31.41+0.31} contains a high-mass hot core with an approximate
mass of 25\,M$_\odot$ \citep{oso09}. Observational estimates of
glycolaldehyde fractional abundance give
$x$(CH$_2$OHCHO)$\sim$10$^{-8\pm2}$ \citep[Beltr{\'a}n, priv. comm.,
based on ][]{bel09}. We therefore use this as a benchmark by which to
judge whether the chemical pathways we test are sufficiently
productive in context. We have explored a number of parameters through
chemical models, all of which are detailed in Table~\ref{tab:models}.

\subsubsection{The efficiency of gas-phase routes to glycolaldehyde formation in hot cores}

Initially we tested whether there was sufficient HCO formed in the gas
phase to make reactions~\ref{HCOdimerst}--\ref{HCOdimerend} an
efficient pathway to the production of glycolaldehyde (see models 1--3
in Table~\ref{tab:models}).  To this end, we disabled the
grain-surface formation of HCO from CO, so that the only pathway for
glycolaldehyde formation was through
reactions~\ref{HCOdimerst}--\ref{HCOdimerend} using gas-phase HCO,
both in the gas phase and the solid phase via freeze-out. This
resulted in no formation of glycolaldehyde after Phase {\sc i}, and
only traces of glycolaldehyde after Phase {\sc ii}. We repeated this
test, but without an energy barrier in the grain-surface hydrogenation
of HOCCOH (reaction \ref{HCOdimermid}; i.e., simulating the tunnelling
of the adsorbed H atom through the energy barrier). This resulted in
the formation of small amounts of glycolaldehyde,
$x$(CH$_2$OHCHO)$\simeq$2$\times$10$^{-12}$\ (model 2). Reducing the
gas-phase barrier had a negligible effect on the abundance of
glycolaldehyde (model 3). From model 1 we see that a very small amount
of glycolaldehyde is formed in Phase {\sc ii}; this essentially tells
us about the efficiency of the gas-phase formation route for
glycolaldehyde.  Using the adopted rate coefficients for reactions
\ref{HCOdimermid} and \ref{HCOdimerend}, $x$(CH$_2$OHCHO) =
$\sim$10$^{-18}$\ is formed, an amount that is insignificant compared
to grain-surface formation mechanisms which occur during Phase {\sc
  i}. There is a dependence on the values of these rate coefficients:
for instance, increasing the rate coefficients for these two
hydrogenations from 10$^{-14}$\ to
10$^{-10}$\,cm$^3$\,s$^{-1}$\ (which is an exceptionally large value
for such a gas-phase reaction at 10\,K) results only in a fractional
abundance of $x$(CH$_2$OHCHO)$\sim$10$^{-10}$. Since this is at the
very lower limit of our observational range, we may conclude, as in
\citet{woo12}, that a gas-phase production method of glycolaldehyde is
unlikely.

\subsubsection{The efficiency of grain-surface routes to glycolaldehyde formation in hot cores}

Given the lack of HCO in the gas phase, we investigated a
grain-surface formation route for glycolaldehyde in more
detail. Adopting a conversion efficiency of 25\% for the production of
HCO from freezing-out CO\footnote[2]{The remainder becomes CH$_3$OH
  (5\%), H$_2$CO (10\%) or stays unhydrogenated, as CO (60\%)}, and
again presuming that reaction \ref{HCOdimermid} is barrierless due to
the tunnelling of H, we considered several more models (models 4--9 in
Table~\ref{tab:models}). With a standard collapse time of
$\sim$5\,Myr, we attain a fractional abundance of glycolaldehyde of
8$\times$10$^{-6}$ (model 4), several orders of magnitude larger than
with no grain-surface hydrogenation of CO to HCO (model 2). This final
abundance is also sufficient to meet our observationally-constrained
criterion for the abundance of glycolaldehyde in G31.41+0.31.

Glycolaldehyde is increased significantly in abundance if we allow the
collapsed core (with a representative density of 10$^7$\,cm$^{-3}$) to
persist for another 5\,Myr without further collapse. In this case, the
abundance of glycolaldehyde reaches 1.3$\times$10$^{-5}$\ (model 5),
an increase of a factor of $\sim$2 when the collapse timescale is
doubled. Since observationally it is not easy to pinpoint
\textquotedblleft $t$=0\textquotedblright, collapse timescales cannot
be accurately assessed. From our modelling, we see that the longer the
chemistry is allowed to evolve in Phase {\sc i}, the more beneficial
it is for the development of complex chemistry.

Models 6 and 7 test the impact of non-thermal desorption mechanisms on
the development. The mechanisms themselves are described in
\citet{vit04} and \citet{rob07}. They have the effect of slightly
increasing the abundance of glycolaldehyde, since they remove CO from
the grain surfaces, which can then re-accrete to generate more
adsorbed HCO.

Finally, we adjusted the sticking coefficient \citep[the
  \textquoteleft{\it fr}\textquoteright\ parameter of][]{vit04}, but
for all reasonable values of this variable ({\it fr} = 0.05--0.2,
where our standard value was {\it fr} = 0.10), the final fractional
abundance of glycolaldehyde did not vary beyond the range
4.5--7.4$\times$10$^{-6}$.

\subsubsection{The impact of barrier size in grain-surface routes to glycolaldehyde formation}

Using model 4 as a basis, we investigated whether it was the size of
the reaction barrier or the amount of CO converted to HCO on freezeout
that affected the final abundance of glycolaldehyde, in models
4.1--4.9. Results of these models are shown in
Table~\ref{tab:4models}.

\begin{deluxetable*}{lcccc}
\tabletypesize{\small}
%\rotate
\tablecaption{Summary of model parameters for 25\,M$_\odot$\ models: detailed models \label{tab:4models}}
\tablewidth{0pt}
\tablehead{
\colhead{\#} & \colhead{Grain-surface} & \colhead{CO hydrogenation} & 
\colhead{$x$(CH$_2$OHCHO)} & \colhead{$x$(CH$_2$OHCHO)} \\
\colhead{} & \colhead{barrier (K)} & \colhead{(\%)} & 
\colhead{Phase {\sc i}}& \colhead{Phase {\sc ii}}
}
\startdata
(4) & \ldots & 60\% CO, 25\% HCO, 10\% H$_2$CO, 5\% CH$_3$OH & 7.9$\times$10$^{-6\phantom{1}}$ & 7.9$\times$10$^{-6\phantom{1}}$ \\
\tableline
4.1 & 1\,108 & 60\% CO, 25\% HCO, 10\% H$_2$CO, 5\% CH$_3$OH & negligible & 2.0$\times$10$^{-14}$\\
4.2 & \phantom{1\,}554 & 60\% CO, 25\% HCO, 10\% H$_2$CO, 5\% CH$_3$OH & negligible & 2.0$\times$10$^{-14}$ \\
4.3 & \phantom{1\,}277 & 60\% CO, 25\% HCO, 10\% H$_2$CO, 5\% CH$_3$OH & 1.4$\times$10$^{-17}$ & 2.0$\times$10$^{-14}$ \\
4.4 & \phantom{1\,}139 & 60\% CO, 25\% HCO, 10\% H$_2$CO, 5\% CH$_3$OH & 1.5$\times$10$^{-11}$ & 1.5$\times$10$^{-11}$ \\
4.5 & \phantom{1\,0}70 & 60\% CO, 25\% HCO, 10\% H$_2$CO, 5\% CH$_3$OH & 1.3$\times$10$^{-8\phantom{1}}$ & 1.3$\times$10$^{-8\phantom{1}}$ \\
\tableline
4.6 & \ldots & 60\% CO, 10\% HCO, 20\% H$_2$CO, 10\% CH$_3$OH & 2.9$\times$10$^{-6}$ & 2.9$\times$10$^{-6}$ \\
4.7 & \ldots & 60\% CO, 5\% HCO, 22.5\% H$_2$CO, 12.5\% CH$_3$OH & 1.3$\times$10$^{-6}$ & 1.3$\times$10$^{-6}$  \\
4.8 & \ldots & 60\% CO, 1\% HCO, 24.5\% H$_2$CO, 14.5\% CH$_3$OH & 1.5$\times$10$^{-7}$ & 1.5$\times$10$^{-7}$  \\
4.9 & \ldots & 60\% CO, 0.1\% HCO, 25\% H$_2$CO, 14.9\% CH$_3$OH & 2.9$\times$10$^{-9}$ & 2.9$\times$10$^{-9}$  
\enddata
\tablecomments{
Reaction energy barriers refer to reaction~\ref{HCOdimermid}. The
conversion between CO and HCO happens via grain-surface hydrogenation
upon freeze-out (column 3; see text.)  Glycolaldehyde abundances are
given in the solid phase for Phase {\sc i} and gas phase for Phase
{\sc ii}.}
\end{deluxetable*}

From models 4.1--4.5 we see that the barrier height in
reaction~\ref{HCOdimermid} has a significant effect, with only small
amounts of glycolaldehyde being formed if the barrier is significantly
larger than $\sim$150\,K. Once the energy barrier drops below
$\sim$100\,K, observational estimates of glycolaldehyde in
\object{G31.41+0.41} are met, with a 25\% conversion from CO to HCO.

\subsubsection{The impact of CO $\longrightarrow$ HCO conversion in grain-surface routes to glycolaldehyde formation}

Given that the reaction energy barrier in reaction~\ref{HCOdimermid}
is reduced sufficiently by H-atom tunnelling, the amount of CO
converted to HCO following freeze-out becomes important. If this
conversion is very efficient (25\%, model 4), the glycolaldehyde
fractional abundance can be as large as 8$\times$10$^{-6}$. However,
even with a low conversion efficiency of 0.1\%, enough glycolaldehyde
is produced to match our observational constraints (model 4.9 in
Table~\ref{tab:4models}).

\subsection{Modelling a low-mass (1\,M$_\odot$) core}

\begin{deluxetable*}{lcccccc}
%\tabletypesize{\small}
%\rotate
\tablecaption{Summary of model parameters for 1\,M$_\odot$\ models. \label{tab:lmmodels}}
\tablewidth{0pt}
\tablehead{
\colhead{\#} & \colhead{Gas-phase} & \colhead{Grain-surface} & \colhead{CO
  $\rightarrow$ HCO} & \colhead{$fr$} &
\colhead{$x$(CH$_2$OHCHO)} & \colhead{$x$(CH$_2$OHCHO)} \\ \colhead{} & \colhead{barrier (K)} & \colhead{barrier
  (K)} & \colhead{(\%)} & \colhead{} & \colhead{Phase {\sc i}}& \colhead{Phase {\sc ii}}
}
\startdata
10 & 1\,312 & 1\,108 & \ldots & 0.012 & negligible & 3.2$\times$10$^{-21}$\\
11 & 1\,312 & \ldots & \ldots & 0.012 & 7.2$\times$10$^{-14}$ & 7.2$\times$10$^{-14}$  \\
12 & 1\,312 & 1\,108 & 25\phantom{.1} & 0.012 & negligible & 5.4$\times$10$^{-15}$ \\
13 & 1\,312 & \ldots & 25\phantom{.1} & 0.012 & 1.3$\times$10$^{-5\phantom{1}}$ & 1.3$\times$10$^{-5\phantom{1}}$ \\
14 & 1\,312 & \ldots & \phantom{2}1\phantom{.1} & 0.012 & 4.4$\times$10$^{-7\phantom{1}}$ & 4.4$\times$10$^{-7\phantom{1}}$ \\
15 & 1\,312 & \ldots & \phantom{2}0.1 & 0.012 & 2.4$\times$10$^{-8\phantom{1}}$ & 2.4$\times$10$^{-8\phantom{1}}$ 
\enddata
\tablecomments{ Reaction energy barriers refer to
  reaction~\ref{HCOdimermid}. $n_\mathrm{fin}$ for all models is
10$^8$\,cm$^{-3}$. The conversion between CO and HCO
  happens via grain-surface hydrogenation upon freeze-out (column 4;
  see text.)  Glycolaldehyde abundances are given in the solid phase
  for Phase {\sc i} and gas phase for Phase {\sc ii}.}
\end{deluxetable*}

Since glycolaldehyde has recently been detected in the low-mass binary
protostellar system, \object{IRAS16293-2422} \citep{jor12}, towards
both protostars, we have modelled a low-mass core with a nominal mass
of 1\,M$_\odot$. We use the time-dependent temperature profile of
\citet{awa10}:
\begin{equation}
  T = 10 + (0.1927\times t^{0.5339})\quad\mathrm{K},
\end{equation}
(for core age, $t$) which fits the empirical data of \citet{sch02}
well.

\citet{jor12} estimate that the fractional abundance of glycolaldehyde
towards \object{IRAS16293-2422} is 6$\times$10$^{-9}$, which adds a
constraint to our modelling. Results of selected models are shown in
Table~\ref{tab:lmmodels}. We see a similar result to that in the
high-mass case, that with a reduction of the energy barrier in the
grain-surface hydrogenation of HOCCOH (via the tunnelling of an H
atom, for example), sufficient glycolaldehyde can be formed to match
observational estimates for \object{IRAS16293-2422}. We stress that
the calculated abundances are upper limits, since we do not consider
destruction of glycolaldehyde, and neither do we consider alternative
reaction channels for HCO, for example.

Models 10 and 12 indicate that a reaction barrier of 1\,108\,K on the
surface is prohibitive for the formation of glycolaldehyde via our
suggested mechanism. Removal of this barrier means that glycolaldehyde
formation becomes more efficient, although the efficiency is limited
by the amount of HCO on grain surfaces. Even a small amount (0.1\%) of
conversion of CO to HCO via grain-surface hydrogenation is sufficient
to meet the observationally-derived estimate for
\object{IRAS16293-2422} (model 15). The fact that the final abundance
of glycolaldehyde produced in the model is governed by the amount
formed in Phase {\sc i} shows that gas-phase formation of
glycolaldehyde is inefficient, even at the higher density of the
low-mass core model.

\section{Discussion and conclusions}

The detection of COMs in recent years, particularly in star-forming
regions, gives a hint of the molecular complexity of the local
universe that awaits discovery. The standard approaches of chemical
modelling or performing laboratory experiments to understand and
explain the presence of these COMs are powerful, but more so when
combined together. Here we have combined quantum chemical calculations
on the energetics of a particular formation route with the modelling
of a large network of chemical reactions. From a reaction chemistry
perspective, the proposed formation mechanism looks promising, but an
investigation into the potentially limiting parameters (e.g., the
availability of HCO) is necessary to test its viability.

As in our previous work \citep{woo12}, we were able to exclude a
gas-phase formation mechanism for glycolaldehyde. Gas-phase formation
only becomes effective once the temperature of the core reaches
$\sim$100\,K, and its yield is dependent on the available resources of
gas-phase HCO. The gas-phase abundance of HCO towards the end of Phase
{\sc ii} of the model matches the observational determination of HCO
abundance in core \object{B1-b} well
\citep[$\sim$10$^{-11}$,][]{cer12}. However, at earlier times, when
glycolaldehyde is forming, the gas-phase abundance of HCO in the model
is very low. The grain-surface pathway proposed in reactions
\ref{HCOdimerst}--\ref{HCOdimerend} produces significant amounts of
glycolaldehyde ($x$(CH$_2$OHCHO) $<$ 8$\times$10$^{-6}$) at lower
temperatures, under the conditions of: i) {\it a grain-surface
  reaction barrier for reaction~\ref{HCOdimermid} of less than
  $\sim$100\,K} and ii) {\it a conversion efficiency of grain-surface
  CO to HCO greater than 0.1\%}. Our estimate of the barrier in
reaction~\ref{HCOdimermid} is 1\,108\,K, which is likely a lower limit
according to CCSD(T) calculations. However, the barrier can be
tunnelled through by H atoms, and so effectively it will be
significantly lower than our estimate. Unfortunately estimates of the
height of the barrier for tunnelling are computationally expensive,
and would take an unfeasibly long amount of time to calculate.

The second condition is also likely, since observed ratios of
CH$_3$OH:CO can be $\sim$1:2 \citep[e.g.,][]{whi11}, significantly
larger than 0.1\%. Since CH$_3$OH is thought to form exclusively
through grain-surface hydrogenation of CO in hot cores, the ratio of
CH$_3$OH:CO gives us some idea of the ratio of HCO:CO on grain
surfaces.

A further effect which may boost the abundance of glycolaldehyde and
other COMs is that of time. Longer collapse timescales mean more time
available for complex molecule formation.

In conclusion, fractional abundances of glycolaldehyde which match the
observed estimates in hot molecular core \object{G31.41+0.31} and
low-mass binary protostar \object{IRAS 16293-2422} are attainable
through a previously undiscovered formation mechanism which we have
investigated and quantified using combined techniques of quantum
chemical calculations and chemical modelling. The efficiency of the
mechanism relies upon there being a small ($\lesssim$100\,K)
activation energy barrier, including tunnelling considerations, for
the reaction HOCCOH + H $\longrightarrow$ OCH$_2$CHO and the
availability of adsorbed HCO, both of which are to be expected.

\begin{figure}
\epsscale{1.00}
\plotone{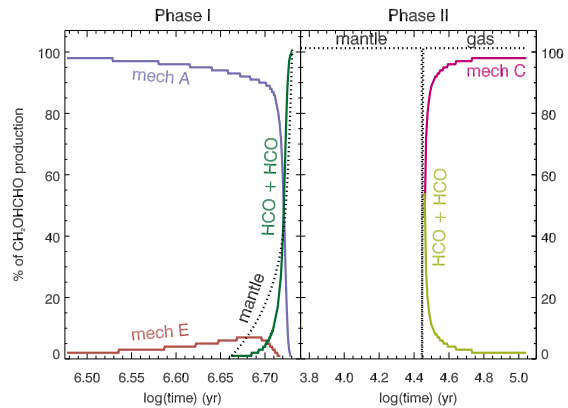}
\caption{Contributions to the production of glycolaldehyde in Phase
  {\sc i} and Phase {\sc ii}. Mechanisms A--E refer to designations in
  \citet{woo12}. The green curves labelled \textquotedblleft HCO +
  HCO\textquotedblright\ refer to the pathway discussed in the
  current paper, and shows the results derived from model 4
  (Table~\ref{tab:models}). Black lines show for illustration the
  change in abundance of glycolaldehyde with time, in the solid phase
  and gas phase (labelled \textquotedblleft
  mantle\textquotedblright\ and \textquotedblleft
  gas\textquotedblright\ respectively.) The scale is not shown, but ranges
  from log(x(CH$_2$OHCHO)) = -14\ldots-5 over the vertical
  axis.\label{fig:comparison}}
\end{figure}

This mechanism adds to our understanding of potential formation routes
of glycolaldehyde. Our previously-favoured mechanisms for the
formation of glycolaldehyde were: \setcounter{equation}{0}
\begin{eqnarray}
\rm A.\:CH_3OH + HCO &\longrightarrow& \rm CH_2OHCHO + H \\
\rm D.\:CH_2OH  + HCO &\longrightarrow& \rm CH_2OHCHO.
\end{eqnarray}
We have added a new potential route,
\begin{eqnarray}
\nonumber\rm 2HCO + 2H &\longrightarrow& \rm CH_2OHCHO.
\end{eqnarray}
As an illustrative exercise, we have run a model with all six
formation pathways for glycolaldehyde in effect: the five mechanisms
from \citet{woo12}, and the dimerisation of HCO. We use the
\textquotedblleft standard rates\textquotedblright\ for mechanisms
A--E, and the model 4 parameters for the model in general.  In
Fig.~\ref{fig:comparison} we show which mechanisms dominate the
formation of glycolaldehyde. As in \citet{woo12}, mechanism A
dominates at early times in Phase I, when there is little
glycolaldehyde formed. The dimerisation mechanism is very effective at
late times, and produces the bulk of glycolaldehyde in this case. It
replaces mechanism D as most efficient, i.e., HCO + HCO dominates over
CH$_2$OH + HCO. HCO is typically $\sim$100 times more abundant than
CH$_2$OH in the model, with a 25\% CO $\rightarrow$ HCO conversion
rate. In Phase II, glycolaldehyde is mostly solid until the
temperature of the core rises to $\sim$100\,K. There is no
grain-surface chemistry in Phase II of our model, so glycolaldehyde is
not formed. Mechanism C is responsible for the majority of production
once the ices are desorbed from the grain mantle, but as described
above, gas-phase formation of glycolaldehyde only occurs at very low
levels, even at warm ($>$100\,K) temperatures.

\acknowledgments

The authors would like to thank T.~P.~M. Goumans for valuable
contributions to this paper.  Funding for this work was provided by the
Leverhulme Trust to P.M.W. and D.J.B. Z.R. and B.S. wish to
acknowledge access to the HECToR facility though their membership of
the UK’s HPC Materials Chemistry Consortium (MCC). The MCC
is funded by EPSRC (EP/F067496). S.V. and W.A.B. acknowledge the
support from the European Community's Seventh Framework Programme
FP7/2007-2013 under grant agreement no. 238258.

\clearpage

\end{document}